\documentclass[12pt, preprint]{aastex}
\def\ltsima{$\;\buildrel < \over \sim \;$}
\def\simlt{\lower.5ex \hbox{\ltsima}}
\def\gtsima{$\;\buildrel > \over \sim \;$}
\def\simgt{\lower.5ex \hbox{\gtsima}}

\shorttitle{Spitzer observations of HD}
\shortauthors{Neufeld et al.}

\begin{document}

\title{Spitzer observations of hydrogen deuteride}

\author{David A. Neufeld\altaffilmark{1}, Joel~D.~Green\altaffilmark{2}, 
David~J.~Hollenbach\altaffilmark{3}, Paule~Sonnentrucker\altaffilmark{1},
Gary~J.~Melnick\altaffilmark{4}, Edwin~A.~Bergin\altaffilmark{5}, 
Ronald~L.~ Snell\altaffilmark{6},
William~J.~Forrest\altaffilmark{2},
Dan~M.~Watson\altaffilmark{2}, and Michael~J.~Kaufman\altaffilmark{7}} 

\altaffiltext{1}{Department of Physics and Astronomy, Johns Hopkins University,
3400 North Charles Street, Baltimore, MD 21218}
\altaffiltext{2}{Department of Physics and Astronomy, University of Rochester, Rochester, NY 14627}
\altaffiltext{3}{NASA Ames Research Center, Moffett Field, CA 94035}
\altaffiltext{4}{Harvard-Smithsonian Center for Astrophysics, 60 Garden Street, 
Cambridge, MA 02138}
\altaffiltext{5}{Department of Astronomy, University of Michigan, 825 Dennison Building, Ann Arbor, MI 48109}
\altaffiltext{6}{Department of Astronomy, University of Massachusetts, 710 North Pleasant Street, Amherst, MA 01003}
\altaffiltext{7}{Department of Physics, San Jose State University, 1 Washington Square, San Jose, CA 95192}

\begin{abstract}

We report the detection of interstellar hydrogen deuteride (HD) toward the supernova remnant IC443, and the tentative detection of HD toward the Herbig Haro objects HH54 and HH7 and the star forming region GGD37 (Cepheus A West).  Our detections are based upon spectral line mapping observations of the 
R(3) and R(4) rotational lines of HD, at rest wavelengths of 28.502 and 23.034$\,\mu$m respectively, obtained using the Infrared Spectrograph onboard the {\it Spitzer Space Telescope}.  The HD R(4)/R(3) line intensity ratio promises to be a valuable probe of the gas pressure in regions where it can be observed.
The derived HD/H$_2$ abundance ratios are $(1.19^{+0.35}_{-0.24}) \times 10^{-5}$, 
$(1.80^{+0.54}_{-0.32}) \times 10^{-5}$, and $(1.41^{+0.46}_{-0.33}) \times 10^{-5}$ respectively (68.3$\%$ confidence limits, based upon statistical errors alone) for IC443 (clump C), HH54, and HH7.  If HD is the only significant reservoir of  gas-phase deuterium in these sources, the inferred HD/H$_2$ ratios are all consistent with a gas-phase elemental abundance $[n_{\rm D}/n_{\rm H}]_{\rm gas} \sim 7.5 \times 10^{-6}$, a factor 2 -- 3 below the values obtained previously from observations of atomic deuterium in the local bubble and the Galactic halo.   However, similarly low gas-phase deuterium abundances have been inferred previously for molecular gas clouds in the Orion region, and in atomic clouds along sight-lines within the Galactic disk to stars more distant than 500~pc from the Sun.

\end{abstract}

\keywords{ISM: Molecules --- ISM: Abundances --- ISM: Clouds -- molecular processes}

\section{Introduction}

With a cosmic abundance $> 10^{-5}$, deuterium is more common 
than all but a half-dozen heavy elements. The initial deuterium abundance was 
set by primordial nucleosynthesis (e.g.\ Schramm \& Turner 1998), 
roughly $\rm 100 \, s$ after the Big Bang, 
and has been reduced by stellar nuclear reactions over the subsequent 13.7 Gyr. 
The primordial deuterium 
abundance revealed from intergalactic absorption components (e.g.\ Kirkman et al.\
2003, who obtained [D/[H]=$2.78^{+0.44}_{-0.38} \times 10^{-5}$ toward Q1243+3047)
is in good agreement with models for primordial nucleosynthesis, given 
the baryon densities derived independently from 
an analysis of the cosmic microwave background (e.g.\ Spergel et al.\ 2003). 
Thus, the distribution of 
deuterium in the Galaxy probes both stellar processing and the degree 
to which material is mixed efficiently within the interstellar medium.
The abundance of interstellar deuterium has been measured primarily by 
ultraviolet absorption line observations of atomic deuterium within diffuse 
clouds (reviewed by Moos et al.\ 2002). Recent measurements of the
Galactic deuterium abundance 
reveal significant variations (over the range $\sim 0.7 - 2.2 \times 10^{-5}$)
that are inconsistent with current 
predictions from Galactic chemical evolution models (e.g.\ Wood et al.\ 
2004, Friedman et al.\ 2006, and references therein); understanding the nature of these
inconsistencies is critical to our understanding of chemical evolution within the Galaxy.

While deuterium has been extensively studied in atomic clouds, the deuterium abundance in
molecular clouds is less certain. In diffuse molecular clouds, interstellar hydrogen deuteride
(HD) has been widely observed by means of ultraviolet spectroscopy -- starting over 30 years ago
with the {\it Copernicus} satellite (reviewed by Spitzer \& Jenkins 1975) -- and is the most abundant
deuterium-bearing molecule. However, deriving the deuterium abundance from the HD column in 
diffuse clouds is difficult for several reasons: the chemistry is complex, HD contains only a trace amount of D, 
and the HD abundance is sensitive to the density and UV field in the cloud 
(Lacour et al.\ 2005; Le~Petit et al.\ 2002).

In dense molecular clouds, by contrast, the observations are much fewer.
Prior to the results reported here, 
detections of hydrogen deuteride emissions had been reported toward just one cloud: 
the Orion Molecular Cloud (Wright et al.\ 1999; Howat et al.\ 2002; Bertoldi et al.\ 1999; the latter's
detection of HD from warm shocked gas having been anticipated theoretically by Timmermann 1996).  
In addition, HD R(0) line absorption has been detected toward two
far-IR continuum sources: Sgr B2 (Polehampton et al.\ 2002) and W49 (Caux et al.\ 2002).
HD is not easily detected by infrared emission-line spectroscopy;
its small dipole moment leads to relatively
weak emission and its large rotational constant places its pure rotational transitions at infrared
frequencies that are inaccessible from ground-based telescopes.
On the other hand, a large number of other
deuterium-bearing molecules are readily detected with ground-based radio telescopes. Many
molecules, such as methanol, show high levels of deuterium fractionation, in which the
deuterated/non-deuterated abundance ratio can exceed the cosmic deuterium abundance by more
than four orders of magnitude; several doubly- and even triply-deuterated species have been
observed (Parise et al.\ 2004). 

In this paper, we report a detection of HD in shocked molecular gas
within the supernova remnant IC443, clump C, and tentative
detections in three other sources -- the Herbig-Haro objects HH54 and HH7, and the star-forming
region GGD37 (Cep A W) -- all from observations with the Infrared
Spectrometer (IRS) on the {\it Spitzer Space Telescope}.  The observational results are discussed in \S2, below, and the implied gas density and HD abundance derived in \S3.  The implications of the inferred HD abundance are discussed in \S 4.  

\section{Observations and results}

Two pure rotational lines fall within the 19.5 -- 37.2~$\mu$m
wavelength range covered by the Long-High module on the IRS: the HD R(3) and R(4)
transitions at 28.502 and 23.034 $\mu$m, respectively\footnote{HD R(0) R(1), and R(2) lie at wavelengths inaccessible to the Long-High module on
{\it Spitzer}/IRS, while R(5) and higher transitions are expected to be so weak that upper limits from
Short-High observations fail to provide useful constraints.}.  Weak spectral features have been observed serendipitously at
both wavelengths in several sources, as shown in Figure 1. These plotted spectra were obtained
in spectral line mapping observations of IC443C conducted in GO cycle 1 -- and of HH54, HH7
and GGD37 carried out in the IRAC- and IRS-GTO programs -- along with observations with the
Short-High module. In addition, IC443C, HH54, and HH7 had been observed in the Short-Low
module, to provide complete spectral coverage from 5.2 -- 37 $\mu$m so that the H$_2$ S(0) -- S(7)
transitions -- and the fine structure lines of [FeII], [SI], [NeII] and [SiII] -- could all be mapped.
These other observations, especially those of the H$_2$ pure rotational transitions, tightly constrain
the temperature and column density of the shocked gas, allowing a determination of the HD/H$_2$ abundance ratio.
  
Table 1 summarizes the measured R(3) and R(4) line intensities.   At each of six positions -- 1 each in IC443C, HH7 and GGD37, and 3 in HH54 -- we have obtained averages of all spectra observed within a circular region, the contributing spectra being weighted by a Gaussian taper (HPBW of 15$^{\prime\prime}$) from the center of each such region.  The central positions are given in Table 1.  Full details of the observations toward HH7 and HH54, as well as the data reduction methods we have developed, are given by Neufeld et al.\ (2006).

Searching the NIST database of atomic fine structure lines -- along with the JPL molecular line
list -- for alternative identifications of the 28.502 and 23.034 $\mu$m features, we found no plausible
candidates besides HD. In particular, a large number of water rotational transitions lie in the
wavelength region of interest, but the only such transitions within one spectral resolution element
of 28.502 and 23.034 $\mu$m are very-high-lying transitions that would be accompanied by other --
much stronger -- transitions that are absent in the observed spectra. As a check upon the reality
of the observed spectral features, we have constructed maps of their distribution. Figure 2 shows
the map obtained toward IC443C, which exhibits the highest column density of warm H$_2$ and the
strongest 28.502 and 23.034~$\mu$m features. Here, we compare maps of the 28.502 and 23.034~$\mu$m
features with those of the H$_2$ S(2), which has an upper state energy $E_U/k$ =1682 K, similar to
those of HD R(3) and R(4) ($E_U/k =$ 1271 K and 1895 K).  A map of the H$_2$ S(3) -- for which $E_U/k =$ 2504 K -- is also shown.
The agreement of the observed
morphology in the four lines lends strong support to the identification of HD and eliminates the
possibility that the 28.502 and 23.034~$\mu$m features are (previously unidentified) instrumental artifacts.
In the other sources, the signal-to-noise ratio is insufficient to allow the distribution of the 28.502 and 23.034~$\mu$m features to be mapped reliably; accordingly, we conservatively describe the detection of HD in those sources as tentative.

\section{HD abundances obtained with Spitzer}
Using {\it Spitzer}/IRS observations of the H$_2$ S(0) through S(7) transitions towards these sources, we
can constrain the column density and temperature of the warm, shocked molecular hydrogen very well, 
and can thereby estimate the HD/H$_2$ abundance ratio.  As described by Neufeld et al.\ (2006),
we have fitted the H$_2$ rotational diagrams with a model which invokes two components: a warm component at temperature $T_w \sim 400$~K and a hot component at temperature $T_h \sim 1000$~K.  The temperatures and column densities of the warm and hot gas components are given in Table 1.  Our analysis neglects the effects of subthermal excitation for the H$_2$ transitions that we have observed; while possibly important for the higher-$J$ transitions of H$_2$ (i.e.\ S(6) and S(7)), such effects are negligible for the 
lower-$J$ H$_2$ transitions that probe the most of the gas capable of producing HD R(3) and R(4) emissions.

Unlike the lower-$J$ states of H$_2$, however, the HD level 
populations for J = 5 and 4 are expected to show 
departures from LTE, due to the presence of a small but non-zero dipole moment. 
We used a statistical equilibrium calculation to compute the HD level populations, adopting the collisional rate coefficients of Flower \& Roueff (1999) for 
excitation by H$_2$.
We thereby determined the resultant HD emission, 
given the two-component model parameters derived from our fit to the 
H$_2$ rotational diagram, and assuming (1) that HD is at the same kinetic temperature as H$_2$; (2) that the HD abundance relative to H$_2$, $n({\rm HD})/n({\rm H}_2)$, is the same in both the warm ($\sim$ 400 K) and hot ($\sim$ 1000K) gas components; and (3) that the warm and hot components have the same gas pressure, $p = n({\rm H}_2) T$.  
Typically, the contributions made by the hot and warm components to the observed HD emissions were roughly equal in our excitation model.  The critical densities for R(3) and R(4) -- i.e. the densities at which the departure coefficients for J = 5 and J = 4 equal one-half -- correspond to pressures  $\sim 4 \times 10^7$ and 
$\sim \rm 10^8 \, cm^{-3}\,K$, respectively, so the 
R(4)/R(3) ratio is a useful density indicator for pressures in the 
$\sim 10^7 -\rm 10^8 \, cm^{-3}\,K$ range.  The R(4)/R(3) intensity ratios predicted for IC443C are 
0.32, 0.40, 0.50, 0.61, 0.72 and 0.82 respectively for log$_{10}(n[{\rm H}_2] T/\rm cm^{-3}\,K)$ = 7.0, 7.2, 7.4, 7.6, 7.8, and 8.0.

We varied the gas pressure, $p$, and the HD abundance relative to H$_2$, $n({\rm HD})/n({\rm H}_2)$, to obtain the best fit to the observed HD R(3) and R(4) line strengths.
Because the HD transitions are optically-thin, the HD line intensities scale linearly with $n({\rm HD})/n({\rm H}_2)$.  
%
While our estimates of the gas pressure or density scale in inverse proportion to the assumed collisional rate coefficients, the derived HD abundance would be entirely unaffected by a (uniform) change in the adopted rate coefficients.  Table 1 lists the best-fit gas pressure and HD abundance relative to H$_2$ derived for each region.

In Figure 3, we show the confidence regions for the gas pressure and HD/H$_2$ ratio: 68.3, and 95.4 $\%$ confidence regions are shown for each of the five sources we observed, based upon the statistical errors
on the HD R(3) and R(4) line fluxes and assuming the H$_2$ column densities and temperatures ($N_w$, $T_w$, $N_h$, $T_h$) listed in Table 1. The confidence limits presented here are based solely upon the statistical errors in the measured line fluxes and do not include systematic uncertainties in the HD excitation model.   Such uncertainties are hard to estimate quantitatively, and include possible errors in the collisional rate coefficients and in the assumption that the warm and hot gas components share a common pressure.  The derived HD abundances (although not the gas pressures) appear to be relatively insensitive to the assumed collisional rate coefficients.  If the HD--He (Roueff \& Zeippen 2000) or HD--H rate coefficients 
(Flower \& Roueff 1999) are adopted in place of the HD--H$_2$ rate coefficients, the best-fit HD abundances change by less than $\sim 10\%$.  On the other hand, 
if it is assumed that the warm and hot gas components share a common density rather than being in pressure equilibrium, the best-fit HD abundances increase by up to $\sim 50\%$.


\section{Discussion}

The signal-to-noise ratio is significantly better toward IC443C, HH54FS, and HH7, the only cases for which R(4) is detected at the $5 \, \sigma$ level.  The best fit gas pressures in these sources, found to lie in the range 
$1.6 \times 10^7 - 1.0 \times 10^8 \, \rm cm^{-3}\, K$, correspond to gas densities in the range $4 \times 10^4$ to $2.5 \times 10^5 \, \rm cm^{-3}$ for the warm gas component.  These values are in reasonable agreement with those derived previously for these regions (see discussion of HH7 and HH54 by Neufeld et al.\ 2006; and of IC443 by Snell et al.\ 2005.)  The HD R(4)/R(3) intensity ratio promises to be a valuable probe of the gas pressure in regions where it can be observed.

Our analysis yields best estimates of $(1.19^{+0.35}_{-0.24}) \times 10^{-5}$, 
$(1.80^{+0.54}_{-0.32}) \times 10^{-5}$, and $(1.41^{+0.46}_{-0.33}) \times 10^{-5}$ (68$\%$ confidence intervals) respectively for the HD abundance relative to H$_2$ in IC443C, HH54FS, and HH7, but all three estimates are consistent (within the 68$\%$ confidence intervals) with $n({\rm HD})/n({\rm H}_2) = 1.5 \times 10^{-5}$.  If HD accounts for all the deuterium in the gas phase\footnote{In interpreting their observations of Orion, Bertoldi et al.\ argued that significant destruction of HD occurred in the shocked gas that they observed, as the result of the reaction $\rm H + HD \rightarrow D + H_2$ (Timmermann 1996), and applied a correction in deriving the gas-phase elemental abundance of HD.  In the shocked regions that we have observed, however, the inferred shock velocities ($\sim 10 - 20\, \rm km \,s^{-1}$; Neufeld et al.\ 2006) are significantly smaller than those for the Orion shock; thus the resultant atomic H abundance is expected to be much smaller and the extent of HD destruction negligible.}, this would imply $[n_{\rm D}/n_{\rm H}]_{\rm gas} = 7.5 \times 10^{-6}$, a value which lies a factor 2 -- 3 below those inferred from atomic D absorption line observations of the local bubble ($[n_{\rm D}/n_{\rm H}]_{\rm gas} = (1.52 \pm 0.08)\times 10^{-5}$; Moos et al.\ 2002), atomic D absorption line observations of the halo ($[n_{\rm D}/n_{\rm H}]_{\rm gas} = (2.2 \pm 0.7) \times 10^{-5}$; Sembach et al.\ 2004), and the expected abundance from primordial nucleosynthesis ($[n_{\rm D}/n_{\rm H}]= (2.62^{+0.18}_{-0.20}) \times 10^{-5}$; Spergel et al.\ 2003, given the cosmological parameters determined by the Wilkinson Microwave Anisotropy Probe).  However, it is comparable to those derived for the Orion Molecular 
Cloud from a claimed detection of HD R(5) toward Orion Peak 1 ($[n_{\rm D}/n_{\rm H}]_{\rm gas} = (7.6 \pm 2.9) \times 10^{-6}$; Bertoldi et al.\ 1999) and from
observations of HD R(0) emission from the Orion Bar ($[n_{\rm D}/n_{\rm H}]_{\rm gas} = (1.0 \pm 0.3) \times 10^{-5}$; Wright et al.\ 1999).
It is also remarkably consistent with the deuterium abundances observed recently in atomic clouds along long sight-lines within the Galactic disk 
toward stars further than 500~pc from the Sun and with atomic H column densities in excess of 10$^{20.5}\,\rm cm^{-2}$, which 
yield $[n_{\rm D}/n_{\rm H}]_{\rm gas} = (8.5 \pm 1.0) \times 10^{-6}$ (Hoopes et al.\ 2003; Wood et al.\ 2004).  

The explanation of these variations has been a subject of considerable debate (Friedman et al.\ 2006).  H\'ebrard and Moos (2003) have argued that the lower values inferred for high-$N_{\rm H}$ sight-lines to distant stars represent the typical Galactic elemental abundance for deuterium, requiring a substantial degree of destruction via astration (i.e.\ in stars), while Draine (2004) has argued for the importance of depletion in reducing the gas-phase deuterium abundance, a possibility originally raised by Jura (1982).  In this picture, material within the local bubble is anomalous because it has suffered recent grain destruction in shocks driven by stellar winds or supernovae.
An observed correlation between the gas-phase abundances of deuterium and the refractory element titanium (Prochaska et al.\ 2005), and the possible detection of deuterated polycyclic aromatic hydrocarbons (Peeters et al.\ 2004), have been interpreted as supporting the importance of depletion.
While our determination of the HD abundance in dense shocked clouds does not directly settle this debate, it does suggest that the gas-phase deuterium abundance in dense clouds differs little from that inferred for typical diffuse clouds along high-$N_{\rm H}$ sight-lines to distant stars
within the Galactic disk.

\acknowledgments

This work, which was supported in part by JPL contract 960803 to the Spitzer IRS Instrument Team and by RSA agreement 1263841, is based on observations made with the {\it Spitzer Space Telescope}, which is operated by the Jet Propulsion Laboratory, California Institute of Technology, under a NASA contract.
D.A.N.\ and P.S.\ gratefully acknowledge the additional support of grant NAG5-13114 from NASA's LTSA Research Program.

\vfill\eject
\begin{deluxetable}{lcccccc}
\tabletypesize{\scriptsize}
\rotate
\tablewidth{0cm}
\tablecaption{Observations of HD}  
\tablehead{
Source &GGD37 & IC443C  & HH7 & HH54C & HH54E+K & HH54FS} 
\startdata
R.A. (J2000)  & $\rm 22h\,55m\,55.14s$  
& $\rm 06h\,17m\,42.47s$ 
& $\rm 03h\,29m\,08.42s$ 
& $\rm 12h\,55m\,53.40s$ 
& $\rm 12h\,55m\,54.58s$ 
& $\rm 12h\,55m\,51.34s$ \\
Dec.  (J2000) 
& $\rm +62d\,02^\prime\,02.7^{\prime\prime}$
& $\rm +22d\,21^\prime\,29.1^{\prime\prime}$
& $\rm +31d\,15^\prime\,27.0^{\prime\prime}$
& $\rm -76d\,56^\prime\,04.5^{\prime\prime}$
& $\rm -76d\,56^\prime\,23.5^{\prime\prime}$
& $\rm -76d\,56^\prime\,18.5^{\prime\prime}$ \\
$T_w$ (K)$^a$                      &N/A$^b$ & 457   & 422	& 558	& 470	&  382 	\\
log$_{10} \, (N_w$/cm$^{-2})$$^a$   &N/A & 20.53 & 19.99    	& 19.80	& 19.84	&  19.89\\
$T_h$ (K)$^a$                       &N/A & 1047  & 1029		& 1150	& 1062	&  1025 \\
log$_{10} \, (N_h$/cm$^{-2})$$^a$   &N/A & 20.01 & 19.17	& 18.86	& 19.13	&  19.22\\

R(3) intensity $/\rm \, 10^{-7}\,erg \, cm^{-2} \, s^{-1} \, sr^{-1}$
& $29.0 \pm 2.8$$^c$ & $52.3\pm 3.1$  & $17.9 \pm 2.2$ & $8.0 \pm 1.2$ & $15.6 \pm 2.9$ & $17.7 \pm 1.0$ \\
R(4) intensity $/\rm \, 10^{-7}\,erg \, cm^{-2} \, s^{-1} \, sr^{-1}$
& $13.6 \pm 2.6$& $21.7\pm 3.1$ & $8.7 \pm 1.2$ & $6.2 \pm 1.3$ & $5.8 \pm 1.7$ & $9.3 \pm 1.8$\\
$I({\rm R(4)})/I({\rm R(3)})$ 
& $0.469 \pm 0.100$ & $0.415 \pm 0.064$ & $0.486 \pm 0.092$ & $0.770 \pm 0.199$ & $0.374 \pm 0.130$ & $0.527 \pm 0.107$\\
${\rm log}_{10}(n({\rm H}_2) \rm T/ cm^{-3}\,K)$ (best fit)
& N/A & 7.23 & 7.55 & 7.98 & 7.22 & 7.53 \\
${\rm log}_{10}(n({\rm H}_2) \rm T/ cm^{-3}\,K)$ (68.3$\%$ c.l.) 
& N/A & 7.10 -- 7.36 & 7.37 -- 7.75 & 7.64 -- 8.71 & 6.88 -- 7.51 & 7.32 -- 7.73 \\
${\rm log}_{10}(n({\rm H}_2) \rm T/ cm^{-3}\,K)$ (95.4$\%$ c.l.) 
& N/A & 6.93 -- 7.49 & 7.20 -- 8.03 & $> 7.33$ & 6.53 -- 7.89 & 7.06 -- 7.98 \\
$10^5 \times n({\rm HD})/n({\rm H}_2)$ (best fit) 
& N/A & 1.19 & 1.41 & 0.53 & 2.07 & 1.80 \\ 
$10^5 \times n({\rm HD})/n({\rm H}_2)$ (68.3$\%$ c.l.) 
& N/A & 0.95 -- 1.54 & 1.07 -- 1.87 & 0.41 -- 0.71 & 1.20 -- 4.46 & 1.49 -- 2.34 \\
$10^5 \times n({\rm HD})/n({\rm H}_2)$ (95.4$\%$ c.l.)
& N/A & 0.79 -- 2.23 & 0.82 -- 2.68 & 0.33 -- 1.02 & 0.72 -- 9.93 & 1.27 -- 3.62 \\
\enddata

\tablenotetext{a}{Two-component fit to the H$_2$ line intensities with a warm component at temperature $T_w$ and H$_2$ column density $N_w$,
and a hot component at temperature $T_h$ and H$_2$ column density $N_h$}
\tablenotetext{b}{not available, because the H$_2$ S(3) -- S(7) lines have not been observed}
\tablenotetext{c}{$1 \,\sigma$ statistical uncertainty}
\end{deluxetable}
\clearpage

\begin{figure}
\includegraphics[scale=0.57,angle=-90]{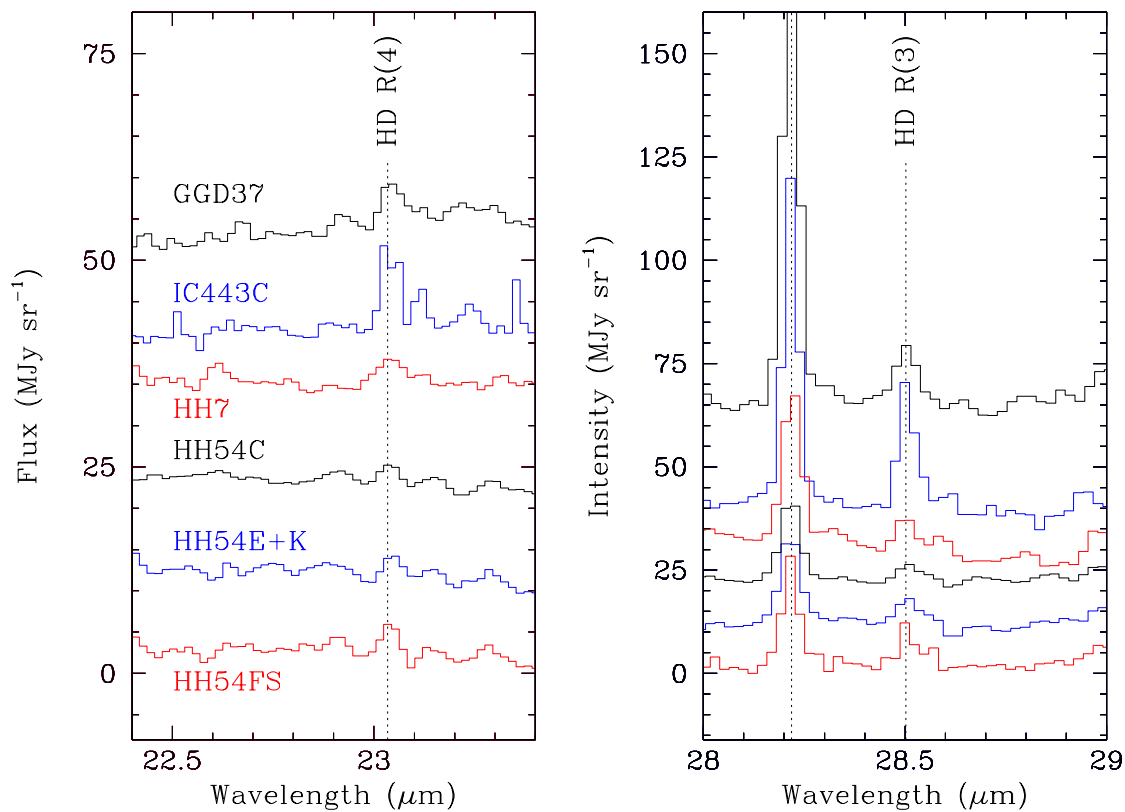}
\figcaption{HD R(3) and R(4) spectra observed for $15^{\prime\prime}$ (HPBW) diameter circular apertures centered at 6 locations defined in Table 1.  From top to bottom the spectra are those of GGD37, IC443C, HH7, HH54C, HH54E+K, and HH54FS.  Arbitrary continuum offsets have been introduced for clarity.
The strong feature at 28.2188~$\mu$m is the S(0) line of H$_2$.}
\end{figure}

\begin{figure}
\includegraphics[scale=1.0,angle=0]{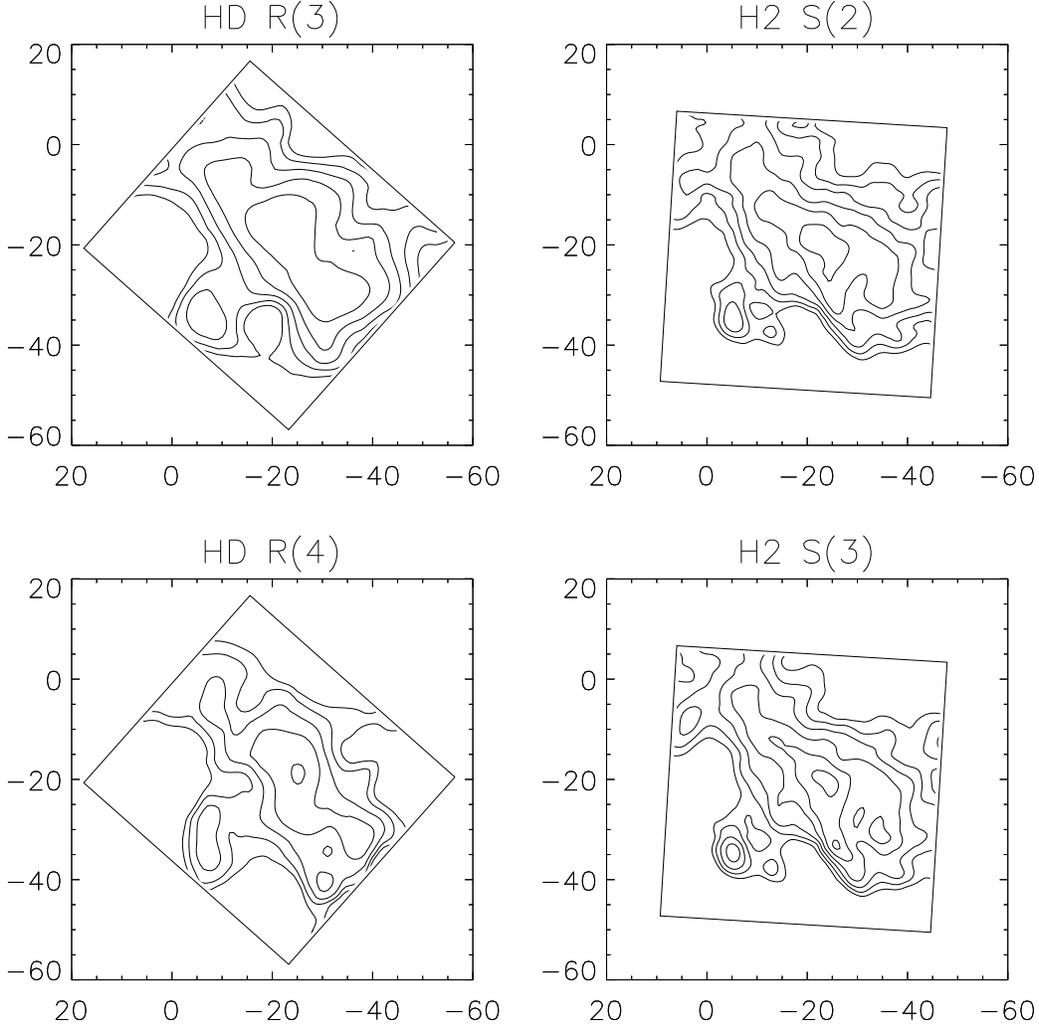}
\figcaption{HD R(3), HD R(4), H$_2$ S(2) and H$_2$ S(3) line intensities observed toward IC443C.  Logarithmic contours are shown, in steps of 0.1 dex, with the lowest contours corresponding to line intensities of $2 \times 10^{-6}$, $1 \times 10^{-6}$, $2 \times 10^{-4}$ and  $1 \times 10^{-3} \rm \, \, erg \, cm^{-2} \, s^{-1} \, sr^{-1}$ respectively for the HD R(3), HD R(4), H$_2$ S(2) and H$_2$ S(3) transitions. The rectangular boxes demark the mapped regions for the Short-Low (H$_2$ lines) and Long-High (HD lines) modules.  The effective spatial resolution is $\sim 3^{\prime\prime}$ and $\sim 10^{\prime\prime}$ (HPBW) respectively for the H$_2$ and HD transitions.    
The horizontal and vertical axes show the R.A. ($\Delta \alpha \rm \cos \delta$) and declination ($\Delta \delta$) offsets in arcsec relative to $\alpha=$06h\,17m\,44.20s, $\delta=22$d\,21$^\prime$49.1$^\prime$$^\prime$ (J2000).  The spectra presented in Figure 1 are averages over a $15^{\prime\prime}$ region centered at offset $(-24^{\prime\prime}, -20^{\prime\prime})$.
}
\end{figure}


\begin{figure}
\includegraphics[scale=0.8,angle=0]{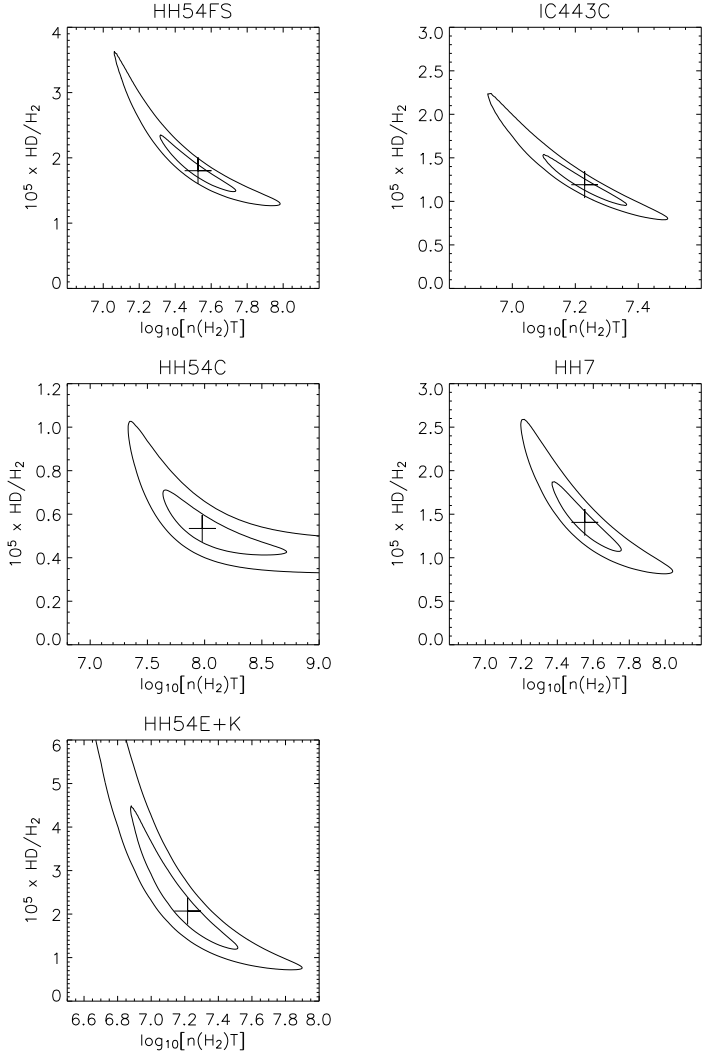}
\figcaption{Contours for $\Delta \chi^2=1$ (inner contour) and 4 (outer contour) as a function of gas pressure, $n({\rm H}_2)T$, and the HD abundance relative to H$_2$, $n({\rm HD})/n({\rm H}_2)$,
are shown for each of the five sources we observed.  These results adopt the H$_2$ column densities and temperatures ($N_w$, $T_w$, $N_h$, $T_h$) listed in Table 1 as priors, and assume that the observational errors in the measured line intensities have a Gaussian distribution function.
The projections of these contours
onto the horizontal and vertical axes yield the 68.3$\%$ (inner contour) and 95.4$\%$ (outer contour) confidence limits on $n({\rm H}_2)T$ and $n({\rm HD})/n({\rm H}_2)$.  The joint probability contents for the inner and outer contours are 39.3 and 86.5$\,\%$ respectively.
Note the different range of gas pressures and HD abundance shown in each panel.  Crosses denote the best-fit values ($\Delta \chi^2=0$).}
\end{figure}

\end{document}